# On the Influence of Carrier Frequency Offset and Sampling Frequency Offset in MIMO-OFDM Systems for Future Digital TV

Youssef Nasser *member IEEE*, Jean-François Hélard *Senior member IEEE*, Matthieu Crussière

*Institute of Electronics and Telecommunications of Rennes, UMR CNRS 6164, Rennes, France*

20 Avenue des Buttes des Coesmes, 35043 Rennes cedex, France

Email : youssef.nasser@insa-rennes.fr

*Abstract*- **This paper investigates the impact of carrier frequency offset (CFO) and sampling frequency offset (SFO) on the performance of different MIMO-OFDM schemes with high spectral efficiency for next generation of terrestrial digital TV. We analyze particularly orthogonal Alamouti scheme, and non-orthogonal (NO) schemes like VBLAST, linear dispersion (LD) code and Golden code. This analysis gives a global view on the best suitable MIMO-OFDM scheme with respect to CFO and SFO. We show that for high spectral efficiency, Alamouti is more sensitive to CFO and SFO. Moreover, we show that all studied MIMO-OFDM schemes are sensitive to CFO when it is greater than 1% of inter-carrier spacing. Their sensitivity due to SFO is less than that due to CFO.**

*Keywords*- **MIMO-OFDM, Carrier Frequency Offset, Sampling Frequency Offset, Iterative receiver.**

## I. INTRODUCTION

Building a next generation digital TV which enables new services such as video contribution for media companies, mobile TV distribution, IPTV distribution becomes a new challenge for broadcasters. Since its inauguration in 1993, digital video broadcast (DVB) project for terrestrial (DVB-T) transmission has fully responded to the objectives of its designers [1]. In 2006, DVB forum launches a study mission to inaugurate original and high defined services for digital TV. During this mission, researches are asked to investigate which technologies could be considered for a second generation terrestrial digital TV called DVB-T2. The main concern of researchers is to support transmission at higher data rates with minimum error probability. It is expected that a multiple input multiple output (MIMO) system combined with orthogonal frequency division multiplexing (OFDM) should take place for that target. However, it is well known that OFDM systems suffer considerably from carrier frequency offset (CFO) and sampling frequency offset (SFO) between transmitter and receiver since CFO and SFO include inter carrier interference (ICI) at the receiving side [2].

This work is carried out within the framework of the European project '*Broadcast for the 21$^{st}$ Century*' (B21C) which constitutes a contribution task force to the consideration engaged by the DVB forum. The main contribution of this work is twofold. First, a generalized framework is proposed for modelling the effect of CFO and SFO on MIMO-OFDM systems. Therefore, we analyze the robustness of different MIMO-OFDM schemes to CFO and SFO using a sub-optimal iterative receiver.

## II. SYSTEM MODEL

We consider in this paper the downlink communication with with two transmit antennas ($M_T$=2) at the base station and two two receiving antennas ($M_R$=2) at the terminal.

Figure 1 depicts the transmitter modules. After channel encoding of information bits $b_k$ with a convolution encoder of coding rate $R$, the encoded bits are mapped according to a Gray quadrature amplitude modulation (QAM) scheme. The Gray mapper assigns $B$ bits for each of the complex constellation points. Using $M_T$ transmitting antennas, a space time (ST) block code (STBC) encoder assigns a ($M_T$,$T$) matrix $X$=[$x_{i,t}$] to each group of $Q$ complex symbols $S$=[$s_1$,…,$s_Q$] from the input of ST module. The ST coding rate is then defined by $L$=$Q$/$T$. The output matrix is transmitted over $M_T$ antennas during $T$ symbol durations i.e. each column is transmitted once during one symbol duration. The different elements $x_{i,t}$ ($i$=1,…,$M_T$; $t$=1,…,$T$) are functions of the input symbols $s_q$ ($q$=1,…,$Q$) depending on STBC encoder type.

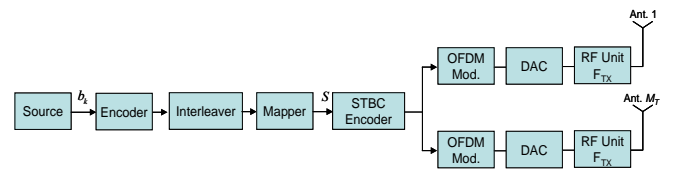

Figure 1- Block diagram of the transmitter.

The symbols at the output of STBC are fed to OFDM modulators of $N$ subcarriers. The output at each OFDM modulator is a sequence of samples having a rate $F_e$=1/$T_e$. After digital to analogue conversion (DAC), the signal is transposed to the transmitter carrier frequency $F_{TX}$ by the RF unit, and transmitted through the channel. At the receiver (Figure 2), it is transposed to base band with the receiver carrier frequency $F_{RX}$ and sampled at sampling frequency $F_s=1/T_s$ using analogue to digital converter (ADC). In this work, we assume equal carrier frequencies $F_{TX}$ (respectively equal sampling frequencies $F_e$) for all transmitting antennas and equal carrier frequencies $F_{RX}$ (equal sampling frequencies $F_s$) for all receiving antennas. The CFO is therefore given by $\Delta F=F_{RX}-F_{TX}$ and the SFO is defined by $1/\Delta T=1/(T_s-T_e)$. After OFDM demodulation, the signal received by the $j^{th}$ antenna at each time sample $t$ on the $n^{th}$ subcarrier could be written as:

$$Y_j[n,t] = \frac{1}{\sqrt{M_T}} \sum_{m=0}^{M_T-1} \sum_{p=0}^{N-1} X_i[p,t] h_{j,i}[p] \phi(n,p) + W_j[n,t] \quad (1)$$

where $h_{j,i}[p]$ is the frequency channel coefficient on the $p^{th}$ subcarrier assumed constant during $T$ OFDM symbols, $W_j[n]$ is the additive white Gaussian noise (AWGN) with zero mean and $N_0/2$ variance. $\phi(n,p)$ is a function of the CFO and SFO, given by:

$$\phi(n,p) = e^{j\pi \frac{N-1}{N}\left(N\Delta FT_s + \frac{\Delta T}{T_s}e(n)+e(n-p)\right)} \times$$
$$\frac{1}{N} \frac{\sin\left(\pi\left(N\Delta FT_s + \frac{\Delta T}{T_s}e(n)+e(n-p)\right)\right)}{\sin\left(\pi\left(N\Delta FT_s + \frac{\Delta T}{T_s}e(n)+e(n-p)\right)/N\right)} \quad (2)$$

with $e(n) = \begin{cases} n & \text{if } n \leq N/2 \\ n-N & \text{elsewhere} \end{cases}$

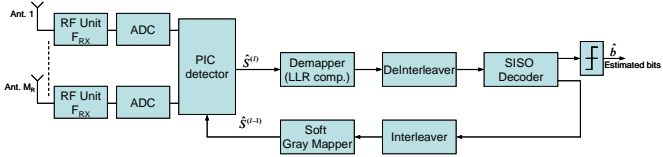

Figure 2- Iterative receiver structure with parallel interference cancellation detector.

The signal received by the $M_R$ antennas on subcarrier $n$ are gathered in a matrix $Y[n]$ of dimension $(M_R,T)$. It can be deduced from (1) by:

$$Y[n] = \phi(n,n)H[n]X[n] + \sum_{\substack{p=1 \\ p \neq n}}^{N} \phi(n,p)H[p]X[p] + W[n]$$
$$= H_{eq}[n]X[n] + \sum_{\substack{p=1 \\ p \neq n}}^{N} \phi(n,p)H[p]X[p] + W[n] \quad (3)$$

In (3), the first term represents useful signal, the second term indicates ICI and the last one is AWGN. $\phi(n,n)$ could be seen as phase rotation and amplitude distortion of the useful signal due to CFO and SFO. The ICI could be seen as an additive noise to the useful signal. It will be neglected in the equalization process. $H[n]$ is a $(M_R,M_T)$ matrix whose components are the channel coefficients $h_{j,i}[n]$. $X[n]$ is a $(M_T,T)$ matrix whose components are the transmitted symbols on the $M_T$ antennas during $T$ OFDM symbols on the $n^{th}$ subcarrier and $W[n]$ is the AWGN.

In order to describe the transmission link with a general model independently of the ST scheme, we introduce the dispersion matrix $F$ such that $X=FS$. Then, we separate the real and imaginary parts of $X[n]$ and $Y[n]$, and we stack them row-wise as done in [3]. We obtain the vector $V[n]$ given by:

$$V[n] = G[n]FS[n] + \sum_{\substack{p=1 \\ p \neq n}}^{N} \phi(n,p)G[p]FS[p] + W[n]$$
$$= G_{eq}[n]S[n] + \sum_{\substack{p=1 \\ p \neq n}}^{N} \phi(n,p)G_{eq}[p]S[p] + W[n] \quad (4)$$

with $G_{eq}[n] = G[n].F$

where $G[n]$ is composed of blocks $G_{j,i}$ ($j=1,\ldots,M_R$; $i=1,\ldots,M_T$) each having ($2T,2T$) elements [3].

In this work, we use an iterative receiver for NO schemes. The estimated symbols $\hat{S}^{(1)}$ at the first iteration are obtained via minimum mean square error (MMSE) filtering as:

$$\hat{s}_u^{(1)}[n] = g_u^{tr}[n]\left(G_{eq}[n].G_{eq}^{tr}[n] + \sigma_w^2 I\right)^{-1} V[n] \quad (5)$$

where $g_u^{tr}[n]$ of dimension (1, $2M_RT$) is the $u^{th}$ column of $G_{eq}$ ($1 \leq u \leq 2Q$). $\hat{s}_u^{(1)}$ is the estimation of the real part ($u$ odd) or imaginary part ($u$ even) of $s_q$ ($1 \leq q \leq Q$) at the first iteration. At each iteration, the demapper provides soft information about transmitted coded bits. The soft information is represented by log likelihood ratios (LLR). After deinterleaving, it is fed to the outer decoder which computes the *a posteriori* extrinsic information of the coded bits. After interleaving, this information will be used by the soft mapper to produce estimation of transmitted QAM symbols. From the second iteration ($l>1$), we perform parallel interference cancellation (PIC) followed by a simple inverse filtering:

$$\hat{V}_u^{(l)} = Y - G_{eq,u}\widetilde{S}_u^{(l-1)}$$
$$\hat{s}_u^{(l)} = \frac{1}{g_u^{tr} g_u} g_u^{tr} \hat{V}_u^{(l)} \quad (6)$$

The exchange of information between detector and decoder runs until the process converges.

III. SIMULATION RESULTS

In this section, we present a comparative study of four MIMO coding schemes: Alamouti, and NO schemes like VBLAST [4], Linear Dispersion (LD) code of Hassibi [5] and Golden code [6]. We use a frequency non selective channel per subcarrier with independent Gaussian distributed coefficients. The performance is computed in terms of bit error rate (BER) versus $E_b/N_0$ ratio for different values of CFO and SFO. CFO is expressed as a function of inter-carrier spacing $1/NT_s$ and SFO is expressed as a function of $N$ and $T_s$ in such a way to guarantee that the SFO does not exceed one sample in one OFDM symbol. The simulations parameters are chosen from those of DVB-T. Figure 3 shows that the sensitivity of Alamouti scheme to CFO for a spectral efficiency $\eta$= 6[b/s/Hz] becomes noticeable for a relative CFO $N\Delta FT_s \geq 1\%$ i.e. $\Delta F \geq 0.01/NT_s$ which is equivalent to 5ppm for 2K mode in DVB-T. Figure 4 shows that SFO introduces degradation for a relative value $N\Delta T/T_s \geq 5\%$. Figure 5 gives the $E_b/N_0$ required to reach a BER=$10^{-3}$ when a CFO occurs at the receiver. Figure 6 shows the effect of SFO at a BER=$10^{-4}$ i.e. it shows the $E_b/N_0$ required to reach a BER=$10^{-4}$. For CFO, we are limited to a measure at a BER=$10^{-3}$ since for Alamouti scheme, a BER floor is obtained. Figures 5 and 6 are given for a spectral efficiency $\eta$= 6[b/s/Hz]. The results for NO schemes are obtained after 3 iterations. These figures show again that all studied MIMO-OFDM schemes are sensitive to CFO for a relative value $N\Delta FT_s \geq 1\%$ and to SFO for a relative value $N\Delta T/T_s \geq 5\%$. Moreover, we can conclude from these figures that the $E_b/N_0$ loss due to CFO is greater than that due to SFO. Indeed, for Alamouti scheme, it is of 9dB for CFO and a BER=$10^{-3}$ and, only 4dB for SFO and a BER=$10^{-4}$. For NO schemes, it is of 2 to 3dB for CFO and 1dB for SFO. Eventually, these figures show that Alamouti scheme is more sensitive to CFO and SFO. This is due to the higher effect of orthogonality loss of Alamouti scheme for higher constellation size.

As a conclusion, when it is based on CFO and SFO, the choice of a given MIMO-OFDM scheme for high spectral efficiency allows us to support non-Alamouti schemes for the second generation of digital TV transmission. However, these schemes require an iterative receiver which is more complex to implement. Moreover, other parameters should be taken into account for the best choice of a MIMO-OFDM scheme.

*Eventually, we note that additional analysis, results and discussions will be available in the final version of the paper.*

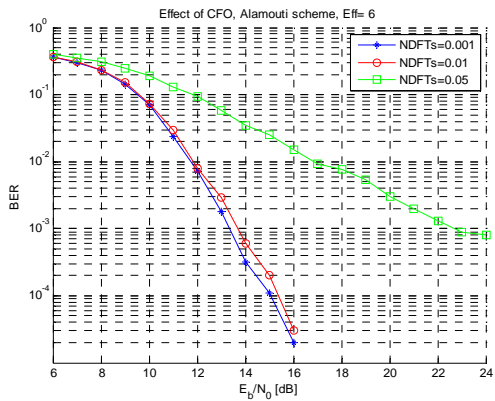

Figure 3- Effect of CFO, Alamouti scheme, Spectral efficiency $\eta$=6 [b/s/Hz] (256-QAM, R=3/4).

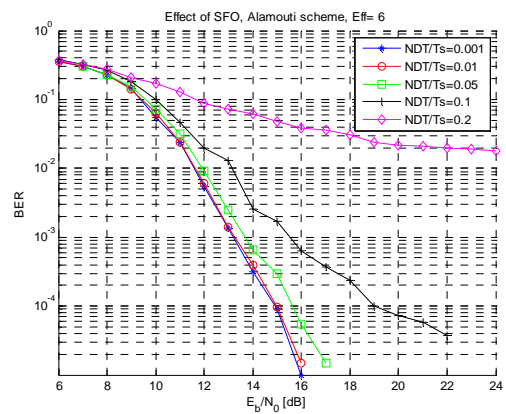

Figure 4- Effect of SFO, Alamouti scheme.

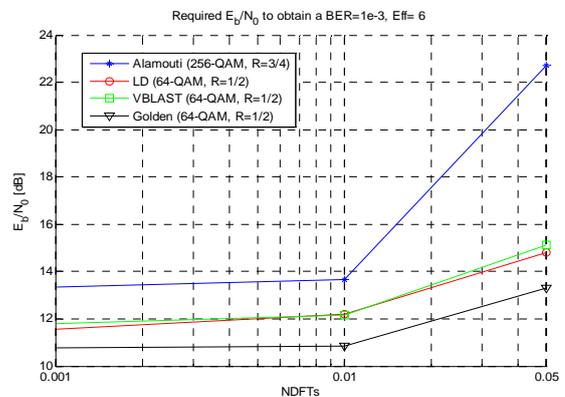

Figure 5- Effect of CFO, Required Eb/N0 to obtain a BER=$10^{-3}$.

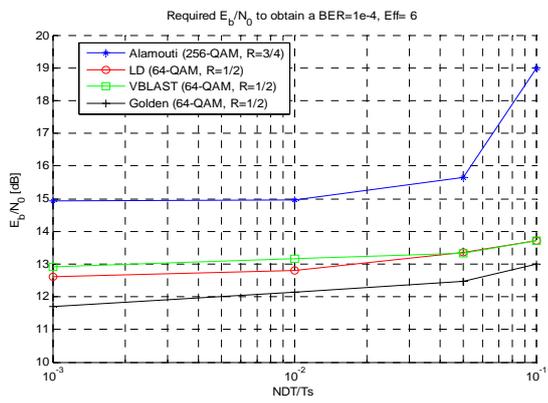

Figure 6- Effect of SFO, Required Eb/N0 to obtain a BER=$10^{-4}$.